\documentclass[journal=jacsat,manuscript=article]{achemso}

\usepackage{graphicx}% Include figure files
\usepackage{dcolumn}% Align table columns on decimal point
\usepackage{bm}
\usepackage{epstopdf}
\usepackage{color}
\usepackage[utf8]{inputenc}
\usepackage[english]{babel}
\usepackage{indentfirst}

\author{O\u{g}uz Y{\i}ld{\i}r{\i}m}
\altaffiliation{Corresponding author}
\email{oguz.yildirim@empa.ch}
\affiliation{Helmholtz-Zentrum Dresden - Rossendorf, Institute of Ion
Beam Physics and Materials Research, Bautzner Landstr. 400, 01328
Dresden, Germany}
\alsoaffiliation{Empa-Swiss Federal Laboratories for Materials Science
and Technology, Ueberlandstr. 129, 8600 D\"{u}bendorf, Switzerland}
\author{Donovan Hilliard}
\email{d.hilliard@hzdr.de}
\affiliation{Helmholtz-Zentrum Dresden - Rossendorf, Institute of Ion
Beam Physics and Materials Research, Bautzner Landstr. 400, 01328
Dresden, Germany}
\alsoaffiliation{Chemnitz University of Technology, Institute of
Physics, Reichenhainer Str. 70, 09126, Chemnitz, Germany}
\altaffiliation{Corresponding author}
\author{Sri Sai Phani Kanth Arekapudi}
\affiliation{Chemnitz University of Technology, Institute of
Physics, Reichenhainer Str. 70, 09126, Chemnitz, Germany}
\author{Ciar\`{a}n Fowley}
\affiliation{Helmholtz-Zentrum Dresden - Rossendorf, Institute of Ion
Beam Physics and Materials Research, Bautzner Landstr. 400, 01328
Dresden, Germany}
\author{Hamza Cansever}
\affiliation{Helmholtz-Zentrum Dresden - Rossendorf, Institute of Ion
Beam Physics and Materials Research, Bautzner Landstr. 400, 01328
Dresden, Germany}
\alsoaffiliation{Dresden University of Technology, Institute of Physics
of Solids, 01062 Dresden, Germany}
\author{Leopold Koch}
\affiliation{Chemnitz University of Technology, Institute of
Physics, Reichenhainer Str. 70, 09126, Chemnitz, Germany}
\author{Lakshmi Ramasubramanian}
\affiliation{Helmholtz-Zentrum Dresden - Rossendorf, Institute of Ion
Beam Physics and Materials Research, Bautzner Landstr. 400, 01328
Dresden, Germany}
\author{Shengqiang Zhou}
\affiliation{Helmholtz-Zentrum Dresden - Rossendorf, Institute of Ion
Beam Physics and Materials Research, Bautzner Landstr. 400, 01328
Dresden, Germany}
\author{Roman B\"{o}ttger}
\affiliation{Helmholtz-Zentrum Dresden - Rossendorf, Institute of Ion
Beam Physics and Materials Research, Bautzner Landstr. 400, 01328
Dresden, Germany}

\author{J\"{u}rgen Lindner}
\affiliation{Helmholtz-Zentrum Dresden - Rossendorf, Institute of Ion
Beam Physics and Materials Research, Bautzner Landstr. 400, 01328
Dresden, Germany}
\author{J\"{u}rgen Fa{\ss}bender}
\affiliation{Helmholtz-Zentrum Dresden - Rossendorf, Institute of Ion
Beam Physics and Materials Research, Bautzner Landstr. 400, 01328
Dresden, Germany}
\alsoaffiliation{Dresden University of Technology, Institute of Physics
of Solids, 01062 Dresden, Germany}
\author{Olav Hellwig}
\affiliation{Helmholtz-Zentrum Dresden - Rossendorf, Institute of Ion
Beam Physics and Materials Research, Bautzner Landstr. 400, 01328
Dresden, Germany}
\alsoaffiliation{Chemnitz University of Technology, Institute of
Physics, Reichenhainer Str. 70, 09126, Chemnitz, Germany}
\author{Alina M. Deac}
\affiliation{Helmholtz-Zentrum Dresden - Rossendorf, Institute of Ion
Beam Physics and Materials Research, Bautzner Landstr. 400, 01328
Dresden, Germany}

\title[An \textsf{achemso} demo]
{Ion-Irradiation-Induced Cobalt/Cobalt Oxide Heterostructures: Printing 3D Interfaces}

\begin{document}

\section{Note}
\textbf{\textcolor{red}{This document is the Accepted Manuscript version of a
Published Work that appeared in final form in ACS Applied
Materials and Interfaces copyright © American Chemical Society
after peer review and technical editing by the publisher. To
access the final edited and published work see https://
pubs.acs.org/articlesonrequest/AOR-rPGr2e7nZ7tzzwK3tnBZ
doi: 10.1021/acsami.9b13503}}

\date{\today}

\begin{abstract}

Interfaces separating ferromagnetic (FM) layers from non-ferromagnetic layers offer unique properties due to spin-orbit coupling and symmetry breaking, yielding effects such as exchange bias, perpendicular magnetic anisotropy, spin-pumping, spin-transfer torques, conversion between charge and spin currents and vice-versa. These interfacial phenomena play crucial roles for magnetic data storage and transfer applications, which require forming FM nano-structures embedded in non-ferromagnetic matrices. Here, we investigate the possiblity of creating such nano-structures by ion-irradiation. We study the effect of lateral confinement on the ion-irradiation-induced reduction of non-magnetic metal oxides (e.g., antiferro- or paramagnetic) to form ferromagnetic metals. Our findings are later exploited to form 3-dimensional magnetic interfaces between Co, CoO and Pt by spatially-selective irradiation of CoO/Pt multilayers. We demonstrate that the mechanical displacement of the O atoms plays a crucial role during the reduction from insulating, non-ferromagnetic cobalt oxides to metallic cobalt. Metallic cobalt yields both perpendicular magnetic anisotropy in the generated Co/Pt nano-structures, and, at low temperatures, exchange bias at vertical interfaces between Co and CoO.  If pushed to the limit of ion-irradiation technology, this approach could, in principle, enable the creation of densely-packed, atomic scale ferromagnetic point-contact spin-torque oscillator (STO) networks, or conductive channels for current-confined-path based current perpendicular-to-plane giant magnetoresistance read-heads.

\end{abstract}

\maketitle
\section{Keywords}  3-D interfaces, magnetic multilayers, perpendicular magnetic anisotropy, exchange bias, ion irradiation

\section{Introduction}
        
 Magnetic interfaces are integral parts of our daily life. The fact that spins of a FM can be tilted\cite{Ota}, biased\cite{Kiwi} or pinned\cite{ZHYuan} by placing them in contact with a non-ferromagnetic material has been exploited in hard disk read heads and other magnetic sensors for decades. More recently, it has been demonstrated that interfaces between ferromagnets and heavy metals are an efficient way to either stabilize magnetic skyrmions at room temperature \cite{Fert, Soum}, or to design energy efficient magnetic storage devices such as race track memory\cite{Parkinrc}. Moreover, the efficiency of spin-transfer torque random access memory (STT-RAM) devices can be dramatically increased by using layers with perpendicular magnetic anisoptropy (PMA), which can be stabilized via the interface between the FM and paramagnetic (PM) layer (e.g., CoFeB/MgO \cite{Pai}, Co/Pt \cite{Ishikawa, MYang, ZLuo, KCai,YCao}, etc.). This list expands as the need for different interface-induced properties increases.
Meanwhile, the ever growing need of smaller, denser and faster communication devices shows no sign of slowing down in the near future. In a conventional manner, the limits of lithographic approaches are pushed to extremes in order to satisfy this demand. An easier path towards ultimate miniaturization can be achieved by locally modifying the material properties by ions\cite{xJFass} or photons\cite{xJEhrler}. Today, ion implantation can be confined to a beam with the dimensions of a single ion\cite{xJDonk}, but this typically requires complicated and expensive implanters which are only capable of performing small area irradiations. Recent studies have reported that simultaneous electric and magnetic patterning of metal oxides by ion-irradiation can be performed by irradiating Co$_3$O$_4$, which is PM at room temperature, through nano-patterned irradiation masks\cite{Dutta, SKimNat}. Indeed, it was demonstrated that upon proton irradiation, Co$_3$O$_4$ reduces to metallic Co. Nevertheless, the underlying physics behind the ion-irradiation-induced oxygen reduction still lacks a consistent understanding. Previous studies claimed that either the implanted proton chemically bonds with oxygen from Co$_3$O$_4$ and forms hydro-oxides, or ballistically removes oxygen from the lattice sites.\cite{Dutta, SKimNat} The ballistic interactions referred to here are the collision events between the energetic protons and the host atoms. Cobalt oxides are good candidates for ion-irradiation-induced patterning, due to their electronic and magnetic properties. Co$_{3}$O$_{4}$ is PM with a semiconducting nature \cite{WChen}, while CoO is an electrically insulating PM above room temperature, and is an antiferromagnet (AFM) below 293 K \cite{FDMorgan}. Therefore, forming conducting and ferromagnetic nanostructures of Co can be easily accomplished by removing oxygen from the lattice by means of energetic ions. 

In this work, we explore the role of using irradiation masks with varying dimensions and it's effect on ion-irradiation-induced patterning. Our results suggest that the ballistic interactions may play a more important role as compared to the chemical reactions which may occur during the ion-irradition-induced reduction of CoO and/or Co$_3$O$_4$  to Co. Furthermore, we also demostrate the formation of multiple CoO/Co and Co/Pt interfaces in a single sample following the spatially confined irradiation of a CoO/Pt multilayer system through a resist mask using broad beam proton irradiation. Our findings are utilized to form 3-dimensional interfaces exhibiting both exchange bias at their vertical Co/CoO interfaces and PMA arising from the horizontal Co/Pt interface\cite{OHellwig, SMaat}.

\section{Mechanism of the ion-irradiation-induced oxide reduction}

6 nm thick CoO and Co$_3$O$_4$ films were grown on silicon substrates using reactive RF magnetron sputtering and capped with 2 nm Pt layers. The chemical composition of the deposited films was controlled by X-ray photoelectron spectroscopy (see the supporting information, S1). Proton (H$^+$) irradiation was performed on extended films, as well as films prepared with irradiation masks. Extended films were irradiated with the following ion doses: 5 $\times$ 10$^{15}$ ions.cm$^{-2}$, 5 $\times$ 10$^{16}$ ions.cm$^{-2}$, 8 $\times$ 10$^{16}$ ions.cm$^{-2}$ and 1 $\times$ 10$^{17}$ ions.cm$^{-2}$, while keeping the ion energy fixed at 0.3 keV for all samples. 

\begin{figure}%[!ht]
\begin{center}
\includegraphics[width=18 cm, clip] {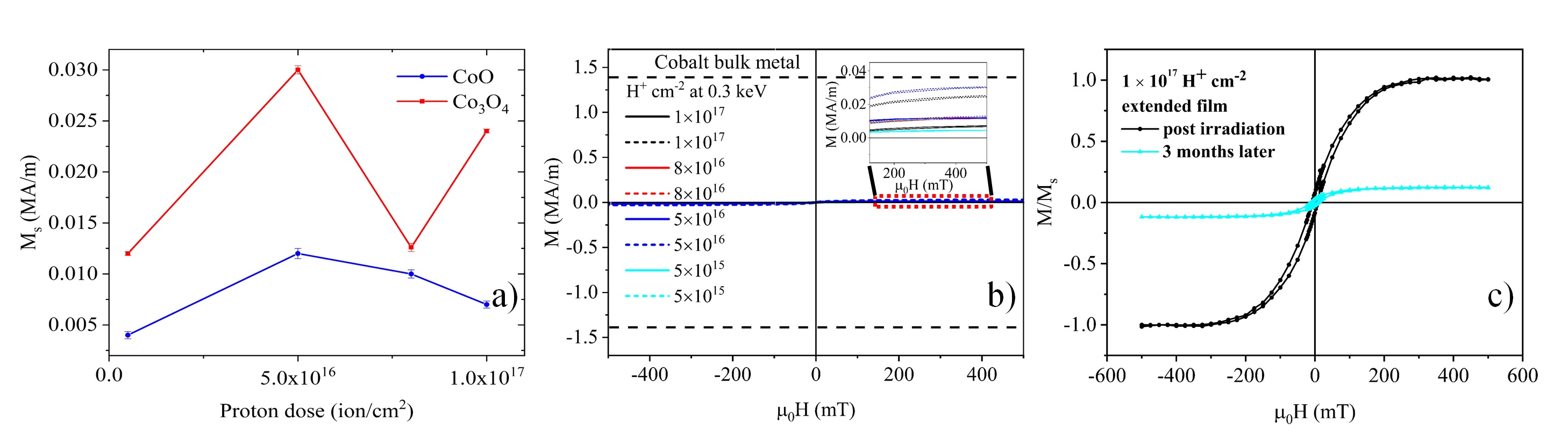}
\caption{\label{fig1} a) Saturation magnetizations of CoO and Co$_{3}$O$_{4}$ as a function of proton irradiation dose obtained at 300K. Values are taken from the corresponding curves given in supporting info S.5 b) Comparison of the magnetization curves to metallic bulk Co at room temperature (dashed line given at around 1.4 MA/m). In the legend, continuous and dashed lines correspond to CoO and Co$_{3}$O$_{4}$, respectively. Inset shows a zoomed-in view. c) A comparison of recovered magnetization post irradiation and again after three months for a proton irradiated extended CoO film at 300K. Normalization is done over initial saturation magnetization value.}
\end{center}
\end{figure}

Figure \ref{fig1}(a) shows the saturation magnetizations ($\emph{M}$$_{S}$) of the irradiated, extended cobalt oxide films as a function of in-plane applied magnetic field. We found that the saturation magnetization values did not have any monotonic dependence on the ion-irradiation dose. Instead, for both oxide phases, the highest $\emph{M}$$_{S}$ was observed for a H$^{+}$ dose of 5 $\times$ 10$^{16}$ ions.cm$^{-2}$. Upon H$^{+}$ irradiation, increased magnetic moment values were observed for all of the samples at room temperature. As compared to CoO, Co$_{3}$O$_{4}$ films show greater magnetization values upon proton irradiation. In all cases the magnetization saturates, albeit gradually, and small coercivities are present. However, as seen in figure 1 (b), the magnetic response of all samples after irradiation is very low when compared to the saturation magnetization of bulk Co metal as the recovered magnetization is found to be about 2\% of the bulk Co metal value.

As mentioned above, the underlying physics behind the proton-irradiation-induced oxygen reduction has not been clearly explained up to now. It has been suggested that either chemical reactions between oxygen and hydrogen or atomic collisions between the incident ions and oxygen atoms are responsible for the observed metallic cobalt formation\cite{SKimNat}. The saturation magnetizations  given in figure \ref{fig1} (a) were measured within five days of irradiation. In order to test the stability of the samples over time, the magnetization measurements were repeated. Magnetization measurements made three months after the initial measurements given in figure \ref{fig1} (c) show a substantial drop in the recovered $\emph{M}$$_{S}$ of up to 90\% indicating reoxidation. Similar reoxidation tendencies were observed in all irradiated, extended films, however, for ease of comparison, the hysteresis loops are shown for only one sample. This indicates that after irradiation, the displaced oxygen atoms stay within the sample in a less stable configuration, consequently allowing them to redistribute over time. It is also worth remembering that all of the samples were capped with Pt layers, thus preventing reoxidation from the atmoshpere. Accordingly, this suggests that oxygen displacement occurs mostly ballistically, i.e., incoming energetic protons transfer their energy to the oxygen atoms leading to displacement of oxygen atoms from the lattice sites and thereby forming metallic cobalt. Results from extended films represent bulk behavior. Yet, in order to exploit this effect in micro/nano devices, it would be necessary to focus the proton beam on selected areas by implementing geometrically patterned irradiation masks.  To this end, we have lithographically patterned striped irradiation masks on top of the CoO and Co$_{3}$O$_{4}$ samples in order to spatially confine the irradiated regions.

\begin{figure*}[!ht]
\begin{center}
\includegraphics[width=18 cm, clip] {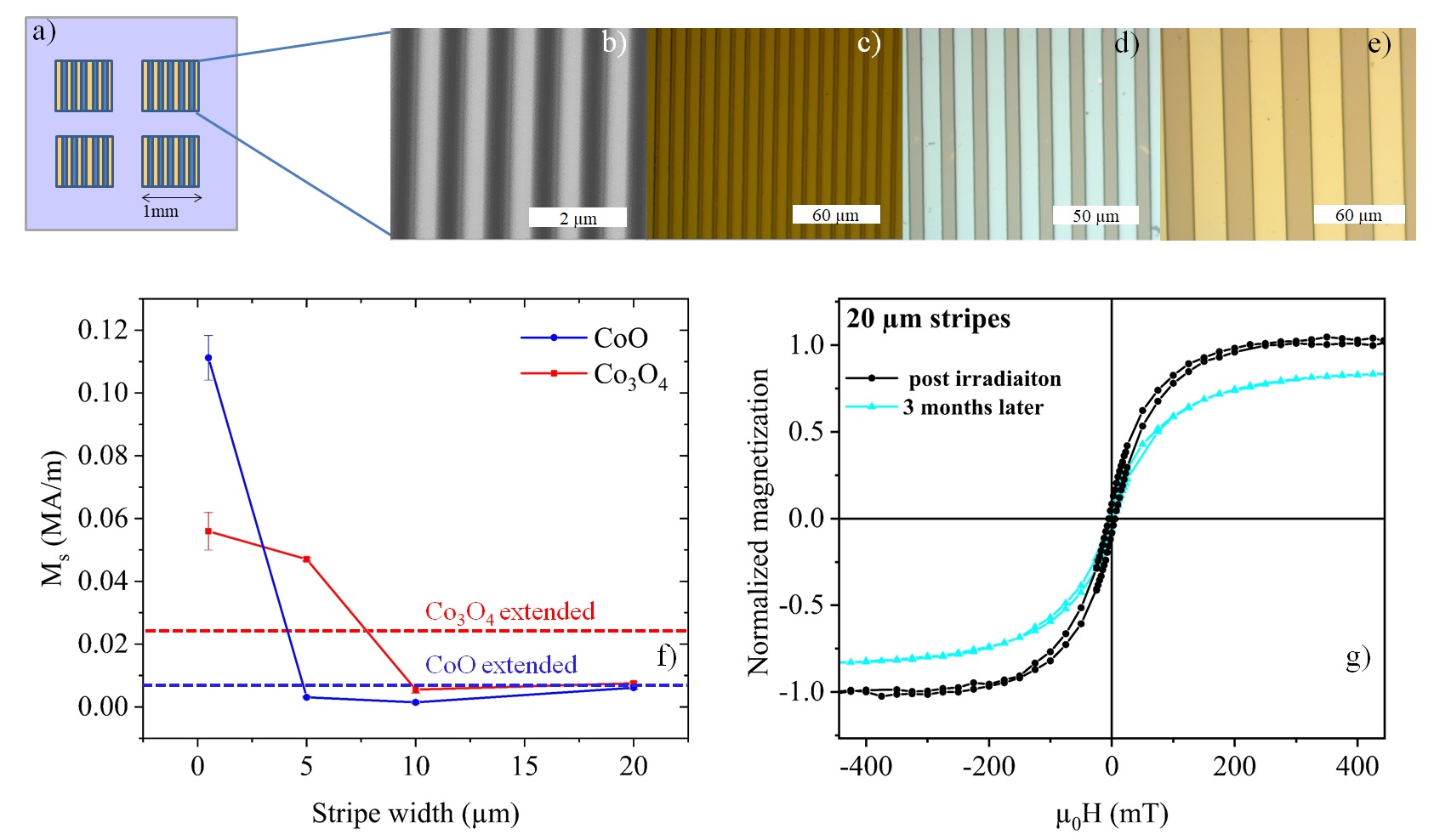}
\caption{\label{fig2} Implementation of the irradiation masks. a) Representation of the location and the dimensions of the masked areas on a film. b) Scanning electron and c-e) optical microscopy images of the irradiation patterns with stripe widths of  500 nm, 5, 10 and 20 ${\mu}$m, respectively. f) Saturation magnetization values at 300K for films irradiated with protons through striped irradiation masks for CoO and Co$_{3}$O$_{4}$. Values are taken from the corresponding curves given in supporting info S.5. Dashed lines show saturation magnetizations obtained from irradiations of extended CoO (blue), Co$_{3}$O$_{4}$ (red) films with the same ion dose. ( g) The time dependence of the magnetic stability after 3 months. Magnetic hysteresis curves of a CoO film at 300K, implemented with a 20 ${\mu}$m wide striped irradiation mask, were recorded post irradiation and then again 3 months later. All of the magnetization curves given here were recorded at room temperature, under an in-plane applied magnetic field which was parallel to the stripe direction.}
\end{center}
\end{figure*}

 Irradiation masks with stripe widths of 500 nm, 5, 10 and 20 ${\mu}$m with a pitch of 1, 10, 20 and 40 ${\mu}$m, were fabricated (figure \ref{fig2} a - e) and the H$^+$ irradiation was performed at an ion fluence of 1 $\times$ 10$^{17}$ ions.cm$^{-2}$.\cite{whydose} Magnetization measurement results after irradiations are shown in figure \ref{fig2} f).  For both oxide types, 500 nm-wide stripes yielded the greatest magnetic response. After irradiation, larger stripes such as 10 and 20 ${\mu}$m, incur a much smaller $\emph{M}$$_{S}$ than that of the narrower stripes. In the CoO case, a slightly different behaviour is observed as seen in figure 2 (f). Indeed, CoO films irradiated through 20, 10, and 5 ${\mu}$m striped irradiation masks show $\emph{M}$$_{S}$ below 0.01 MA/m, while the $\emph{M}$$_{S}$ of the film with 500 nm stripes reaches 0.10 MA/m. Other than the fact that the $\emph{M}$$_{S}$ of this particular sample is one order of magnitude greater, there is no direct dependence on the stripe width seen in films of larger stripes, as is also the case for Co$_{3}$O$_{4}$ (fig. 2 f).  Regardless, it is found that the use of a stripe mask becomes effective when the stripe width is smaller. For CoO, the sample with 500 nm stripes shows greater saturation magnetization as compared to the extended CoO irradiated with the same ion dose. Similarly, for the case of Co$_{3}$O$_{4}$, the effect of the masks become clearer when the stripe width is  5 ${\mu}$m or lower. Stripe widths larger than these resulted in saturation magnetization values comparable to or smaller than the extended films. This could be attributed to the density of the interfaces. 

 Similar to the films irradiated without masks, we repeated the magnetization measurements 3 months later. Magnetization measurements performed 3 months later revealed that the magnetization drops around 20\% (fig. 2 g), which is substantially lower as compared to the films that were exposed to extended irradiations. Dramatic reoxidation of the irradiated extended films as opposed to films where the irradiations are confined to much smaller dimensions cannot be ascribed purely to a possible chemical reduction. In a closed system, capped with Pt, reoxidation can occur only internally (i.e., re-diffusion of displaced oxygen atoms to form cobalt-oxides) and cobalt hydro-oxides are found to be quite stable at room temperature \cite{TDeng}. On the other hand, one can argue that the reoxidation can occur as a result of possible cracking upon proton irradiation. In such a case, complete reoxidation of the all films would be expected, independent of the irradiation mask size. Therefore, the oxygen atoms, removed from the lattice upon proton irradiation, should have a relatively unstable state, which enables them to form cobalt oxide again. On the other side, it is worth mentioning that the above mechanism we propose is one of the possible scenarios that can take place during this process. An investigation focused on finding oxygen locations at the intermediate and final stages of this process could help in explaining this mechanism further.

 \maketitle
\section{Printing perpendicular interfaces}

\begin{figure*}[!ht]
\begin{center}
\includegraphics[width=18 cm, clip] {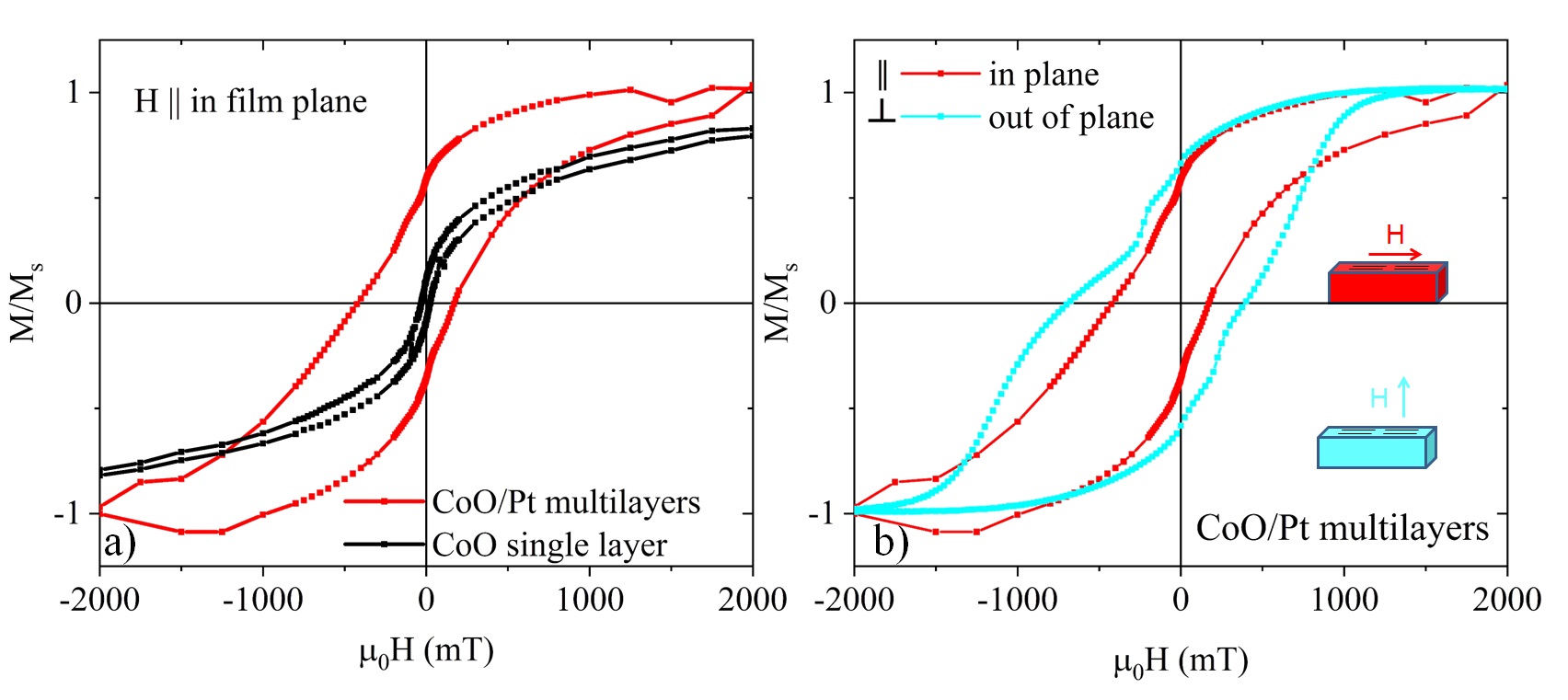}
\caption{\label{fig3} Magnetic hysteresis curves recorded at 10 K after cooling from room temperature under a 7 T applied magnetic field. a) Comparison of irradiated single CoO layer and CoO/Pt multilayers. b) Magnetization curves of multilayered films for different magnetic field directions with respect to the film plane. Normalization was done at a higher field region, however, the image shown here is given for a smaller region due to better visibility. Accordingly, saturation of the single layer differs from 1 in the visible region. This is due to a large PM signal of the single layer which shifts the saturation field to higher fields.}
\end{center}
\end{figure*}

Here, we demonstrate how oxygen reduction by proton irradiation can be exploited in spintronics and how vertical AFM/FM interfaces can be printed. In this study, both a 6 nm thick single layer CoO film and a [CoO(0.8 nm)/Pt(0.8 nm)]$_5$ multilayer film were prepared and capped with a thin Pt cap layer (see methods). Both films were fabricated with 500 nm-wide striped irradiation masks, then irradiated with 0.3 keV protons at an ion dose of 1 $\times$ 10$^{17}$ cm$^{-2}$ (the same conditions given in the previous section). After irradiation, samples cooled down under 7 T magntic field and magnetization curves were recorded at 10 K (figure \ref{fig3}). The magnetic field during cooling and measurement was applied always in the same direction, i.e., for out-of-plane magnetization measurements (figure \ref{fig3} b), the magnetic field was perpendicular to the sample plane during field cooling as well as during the measurements. For in-plane magnetization, the magnetic field was parallel to the sample plane and the stripe direction during cooling and the measurement. Single layer CoO, irradiated through a 500 nm-wide striped mask did not show any exchange bias at 10 K (figure \ref{fig3} a). For a system with well defined Co/CoO interfaces, exchange bias effect under these measurement conditions is expected. On the other hand, the multilayered film, after the same irradiation process through a 500 nm-wide mask, showed a well-defined exchange bias after training, observed as a shift in the field axis of the magnetization curve, as shown in figure (\ref{fig3} a). This indicates that the CoO/Co interfaces in the multilayered film are more defined allowing pinning of the ferromagnetic spins of Co by AFM CoO as compared to the case of single layer CoO. We also found that when the magnetic field is applied out of the film plane, the coercive field increases. However, the exchange bias field decreases slightly from 195 mT to 137 mT at 10 K (figure \ref{fig3} b). The observed difference for different field orientations is consistent with a previous report \cite{SMaat}. This suggests that each individual stripe behaves like an independent film with a well-defined CoO/Co interface showing exchange bias in addition to a Co/Pt interface exhibiting out-of-plane easy axis magnetization.

\begin{figure*}
\begin{center}
\includegraphics[width=17 cm, clip] {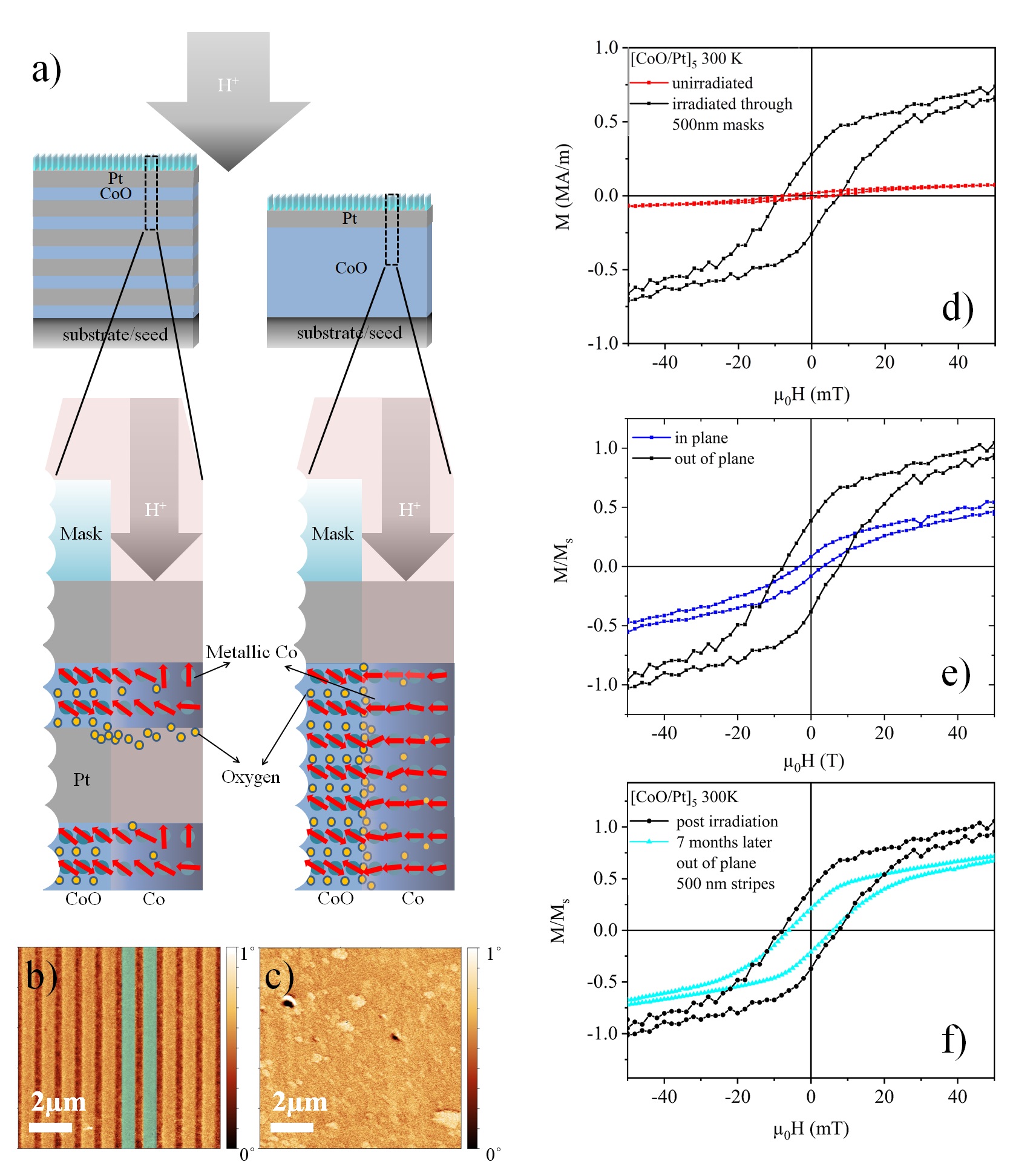}
\caption{\label{fig4n} a) Schematic representation of the evolution of magnetic spins as well as the distribution of the atoms in the proton irradiated CoO/Pt (left) and CoO film (right). b) MFM phase images of irradiated [CoO/Pt]$_{5}$ multilayered and single layered CoO c) films. Bright contrast for b), also highlighted with cyan markers, represent where the mask was placed and the dark contrast is obtained from the irradiated regions. d)  Out-of-plane magnetization curves for unirradiated as well as irradiated [CoO/Pt]$_{5}$ multilayered films. Irradiations were done through 500 nm-wide striped masks. e)  In- (blue) and out-of-plane (black) measurements of the irradiated multilayer sample.  Measurements were performed at 300 K. f) Time dependence of the magnetic stability of the irradiated multilayers after 7 months.}
\end{center}
\end{figure*}

Ferromagnetic behavior accompanied with a relatively large exchange bias point towards a more efficient oxygen removal upon proton irradiation. In addition to the exchange bias occuring at the Co/CoO interface, for a well-defined Co/Pt interface, perpendicular easy-axis magnetization is expected. Indeed, a perpendicular magnetic easy-axis was observed for the irradiated multilayered film at room temperature. The coercive field drops at 300K, yet stays within the typical range for Co/Pt layers (figure 4 e), around a few tens of mT \cite{Landis}. Nevertheless, the shape of the magnetization switch for the irradiated sample is broader (fig. 4 d and e), i.e., canted for a material with a full perpendicular easy axis \cite{Emori} (for comparison to a metallic reference sample, please see the supporting infomration, S4). This canted switching is attributed to the partially damaged Co/Pt interfaces after irradiation (fig. 4 a) and similar behavior was also confirmed by Hall measurements (see the supporting information, S.6). Such damaged interfaces can accomodate Co atoms with in-plane easy-axis magnetization. However, the existence of exchange bias and perpendicular magnetic anisotropy confirms both Co/Pt and CoO/Co interfaces in three dimensions are intact and magnetically effective. 
Formation of the ferromagnetic volumes inside irradiated [CoO/Pt]$_{5}$ multilayered films with 500 nm striped irradiation masks was further confirmed by magnetic force microscopy (MFM) images. Figure 4 b) and 4 c) display room temperature remanence MFM images for [CoO/Pt]$_{5}$ multilayered and  CoO single layered films irradiated through 500 nm striped irradiaton masks, respectively. For the multilayered film, the MFM image shows a ferromagnetic contrast according to the dimensions of the irradiation mask. On the other hand, for the irradiated single layered CoO film, we were not able record any magnetic contrast with MFM, which is most likely due to a very weak magnetic signal leading to a weak stray field and is in line with the magnetization data. In addition, X-ray diffraction (XRD) measurements also indicate the formation of metallic Co after irradiation, seen as a shift of the superstructure peak towards higher diffraction angles (see supplementary material, S2). It is also worth noting that during imaging no crack formation was observed for both samples.

\section{Discussion}

The effect of lateral confinement on the ion-irradiation-induced oxygen reduction is explored. It is found that, for our case, by increasing the number of interfaces either by introducing irradiation masks or by preparing multilayered films, the efficiency of the oxide reduction can be enhanced. Moreover, these interfaces increase the durability of the formed ferromagnetic regions. In comparison to a previous study with Co$_3$O$_4$/Pd multilayers,\cite{SKimNat} the amount of reduced metallic Co is lower in our case using CoO/Pt multilayers. This can be attributed to two different  reasons; (i) Co$_3$O$_4$/Pd is a different system with a different potential to accomodate the excess oxygen compared to a CoO/Pt system, (ii) irradiation mask dimensions. In our case, proton irradiations were done through larger masks compared to that study.\cite{SKimNat} This is of particular importance, because we show that we can increase the efficiency of the oxygen reduction by decreasing the mask`s stripe width. However, it is also worth mentioning that due to lower oxygen reduction efficiency, our results might be describing an intermediate state of proton irradiation-induced oxygen reduction for this particular system. Apart from the oxygen, hydrogen that is implanted inside the material may also be forming some secondary phases as PM cobalt-hydro-oxides. \cite{SKimNat} Yet, these secondary phases must have either a small concentration or are in an amorphous state with smaller grains, such that they are effectively invisible by means of X-rays (see supporting information, S.2). Nevertheless, our results suggest that the mechanism of the irradiation-induced oxygen reduction cannot be purely chemical. Instead, ballistic removal of the oxygen atoms, which requires less energy as compared to the removal of Co or Pt, plays an important role (the threshold proton acceleration energies required to displace Co and O from their lattice position are considered to be 550 and 175 eV, respectively) . \cite{SCShin} It has been shown that, independent of the oxidation mechanism, in the first few monolayers of CoO, oxygen-cobalt bonding is much weaker than in the bulk. \cite{NLWang} Therefore, in our experiments, we expect to have weak bonding between Co and O in the as-grown state, so that oxygen removal would require lower energies. In addition, at temperatures below 373 K, the oxygen diffusion length in Co is around 1 nm. \cite{HGTompkins} Additionally, for Co/Pt multilayers, the lattice mismatch between Co$_{fcc}$ and Pt$_{fcc}$ (${\approx}$ 10\% )\cite{BZhang} induces defect formation at the interfaces. These defects were characterized as vacancies.\cite{JZhang} Due to shorter distances than the diffusion length, Co/Pt interfaces could provide a suitable location for displaced oxygen atoms. This estimation is also in line with a study where it was shown that similar interfaces between Pt and TiO$_2$ could accomodate the free oxygen owing to their higher defect/vacancy concentrations in comparison to the bulk of the layers\cite{Chnauenheim}. So, as a result of its increased vacancy concentration, a Co/Pt interface could be an appropriate location for the removed excess oxygen atoms. Because of the directionality of the proton beam, i.e., the direction of the energy that is transferred to the oxygen atoms, it is expected that more oxygen accumulation occurs in the direction of the ion beam, from the top towards the subtrates direction. This suggests that for each Co layer, the bottom Co/Pt interface may behave like a sink and can be deformed, whereas, the top interface, Pt/Co, can stay intact, giving rise to the observed perpendicular magnetic anisotropy (figure 4 a). In contrast, the bottom Co/Pt interface cannot be the only location where dislocated oxygen atoms are trapped. Because, in such a situation the reduction efficiency would not exceed 50\% (see the supporting information table S.1 and figure S.3). Nevertheless, our results both from single and multilayers show that the density of interfaces either created through irradiation masks or placed during sample preparation plays a crucial role when forming FM structures.

\section{Conclusions and Outlook}

We describe the formation of 3-dimensional interfaces at the nanoscale and address the role of lateral confinement of the irradiated areas. Our results open a new way to form 3-dimensional magnetic and/or conductive heterostructures. Especially, considering how low irradiation energies result in shorter lateral straggle (3nm, see the supporting material S.3) by utilizing this method it is in principal possible to achieve structures with dimensions closer to the ion beam size. On the other hand, reoxidation of the ion-irradiation-reduced Co layers still presents a problem that could limit device lifetime and performance. We believe that further optimization of this method can pave a way for denser networks of spintronic devices in which biasing and perpendicular magnetic anisotropy can be introduced in different dimensions. This method can also be utilized in AFM spintronics where FM/AFM and heavy metal/AFM interfaces are required.\cite{YMHung} Based on our results, further optimization of this process can be done by implementing more interfaces or an appropriate getter/sink layer for oxygen to be absorbed or changing the irradiation ion species could yield sharper interfaces and help improve this method for use in other metal/oxide thin film systems for different applications as well. 

\section{Methods}

CoO and Co$_{3}$O$_{4}$ layers were grown using reactive RF magnetron sputtering. Pt layers were grown under an Ar atmosphere, while oxide layers were grown under an Ar and O$_2$ atmosphere, with varied O$_2$ to Ar ratios (for further details, please see the supporting inormation). Oxide phases were verified using X-ray photoemission spectroscopy. All films were capped with Pt layers on top to protect from oxidation. Films consisting of single CoO or Co$_{3}$O$_{4}$ layers were capped with 2 nm Pt, while multilayers were capped with 2.7 nm Pt. Proton (H$^{+}$) irradiation was performed at 0.3 keV under different doses using active water cooling on the sample holder in order minimize the thermal effects sourced by the irradiation. (see the related section above). All magnetization measurements were done using a Quantum Design magnetic properties measurements system (MPMS-3) equipped with a vibrating sample magnetometer (VSM) head and superconducting quantum intereference device (SQUID) coils. Magnetizaiton measurements were performed using diamagnetic, low-magnetic signal sample holders. Also, some of the SQUID measurements were cross-controlled by measuring with other SQUID-MPMS setups such as MPMS-XL. 500 nm striped irradiation masks were spin coated with 0.3 ${\mu}$m thick ZEP520A positive EBL resist and exposed using electron beam lithography (either Raith150 TWO or Raith eLine).  5 ${\mu}$m, 10 ${\mu}$m and 20 ${\mu}$m striped masks were spin coated with 1.3 ${\mu}$m Shipley S1813 positive photoresist and exposed in a ``Direct Laser Writer''. Resist thicknesses was chosen after series of SRIM simulations in order to make sure that protons would not be able penetrate throught the masks. MFM measurements were performed using a Bruker Multimode setup. Scans were carried out in tapping mode with an interleave lift height of 40 nm.

\section{Associated Content}
\textbf{Supporting Information}

Additional figures;

S1: XPS characterization of as-grown films,

S2: XRD data obtained from as-grown and irradiated CoO/Pt multilayers,

S3: SRIM simulations showing lateral and depth distribution of protons in CoO/Pt multilayer,

S4: Comparison of as-grown and irradiated CoO/Pt sample with a metallic Co/Pt reference sample,

S5: Magnetic field dependent magnetization measurements of single layer cobalt oxide films at room temperature,

S6: Hall measurements for as-grown and irradiated CoO/Pt samples and the metallic Co/Pt reference sample.

Additional table:

Table 1: Summary of saturation magnetization and exchange-bias fields for different samples

\section{Acknowledgements}

This work is supported by the Helmholtz Young Investigator Initiative Grant No. VH-N6-1048. Support of the Nanofabrication Facilities of Rossendorf at the Ion Beam Centre is gratefully acknowledged (Dr. Artur Erbe, Bernd Scheumann). Also, we would like to thank Manuel Monecke, Prof. Dr. Dr. h. c. Dietrich R. T. Zahn and Prof. Dr. Georgeta Salvan from TU Chemnitz for the XPS measurements

\section{Author contributions}

O.Y., D.H., C.F. and A.M.D. designed the experiments. D.H., S. S. P. K. A., H.C., prepared the samples. O.Y., D.H. and S.S.P.K.A. performed the magnetometry experiments. O.Y and  S.S.P.K.A. performed the structural characterization measurements, i.e., XRD and XPS measurements. R.B. conducted ion-irradiations. D.H performed MFM measurements. O.Y. and D.H. wrote the manuscript with the input from all authors.


\begin{thebibliography}{99}
\bibitem{whydose} The reason why this ion dose is chosen is to have comparable data with previously published studies, while the difference between the results obtained from 1 $\times$ 10$^{17}$ ions.cm$^{-2}$ and 5 $\times$ 10$^{16}$ ions.cm$^{-2}$  was not significant.
\end{thebibliography}
\end{document}